\documentclass[11pt]{article}
\usepackage{blois2002,epsfig}
\def\nnbar{$ n \rightarrow \bar{n} \ $}
\def\nnbari{$ \it n \rightarrow \bar{n} \ $}

\def\be{\begin{equation}}
\def\ee{\end{equation}}
\def\bea{\begin{eqnarray}}
\def\eea{\end{eqnarray}}

\begin{document}
\vspace*{4cm}
\title{NEUTRON-ANTINEUTRON OSCILLATIONS}
\author{ Yuri KAMYSHKOV }
\address{Department of Physics, University of Tennessee, Knoxville, TN 37996-1200, USA}

\maketitle
\abstracts{Experimental observation of nucleon instability is 
one of the missing key components required for the explanation of baryon 
asymmetry of the universe. Proton decays with the modes and rates 
predicted by(B$-$L)-conserving schemes of Grand Unification are 
not observed experimentally. There are reasons to believe 
that (B$-$L) might not be conserved in nature, thus leading to the 
nucleon decay into lepton+(X) and to phenomena such as Majorana 
masses of neutrinos, neutrinoless double-beta decays, and most 
spectacularly to the transitions of neutron to anti-neutron. The 
energy scale where (B$-$L) violation takes place cannot be predicted 
by theory and therefore has to be explored by experiments. 
Different experimental approaches to searching for (B$-$L)-violating transition 
of neutron to antineutron are discussed in this paper. Most powerful search for \nnbar 
can be performed in a new reactor-based experiment at HFIR/ORNL where sensitivity 
can be $\sim$ 1,000 times higher than in the previous experiments.}

\section{Introduction}

Our expectations for experimental observation of nucleon instability \cite{Goldhaber}
are based on two outstanding ideas in contemporary physics: the need for 
explanation of baryon asymmetry of the universe (BAU) \cite{Sakharov,Kuzmin} and the concept 
of Unification of particles and forces \cite{Salam,Georgi}. Presence of baryon 
instability in early universe is one of the components required \cite{Sakharov}
for the explanation of BAU. However, BAU as an experimental fact does not tell 
us how baryon instability manifests itself. Particular modes of nucleon decay
are predicted by theoretical schemes beyond the Standard Model (like Unification 
models) that yet have no solid experimental foundation. There is nothing sacred 
in conservation of baryon number. Even within the Standard Model, baryon number 
is not conserved at the non-perturbative level \cite{tHooft}. The latter non-conservation 
is very weak at the present temperature of the universe and does not lead to any 
observable effects. 

In spite of significant experimental attempts nucleon instability (other 
than regular $\beta $-decay) so far has not been discovered \cite{PDG} 
suggesting further experimental efforts with increased mass of the detectors 
\cite{NNN99} as well as the experiments in alternative directions \cite{YK1}. One of 
such possible alternative experiments is a new sensitive search for neutron to 
antineutron transition \cite{YK2} that might explore stability of matter at a 
lifetime scale an order of magnitude beyond the reach of contemporary 
nucleon decay experiments.

\section{Why $n \rightarrow \bar{n}$ ?}

Let us discuss why experimental observation of \nnbar transitions is important. 
In the baryon disappearance the conservation of the angular momentum (nucleon spin 1/2) 
would require at least one fermion to appear in the final state. A nucleon or leptons 
$e$, $\mu $, three neutrinos and their antiparticles are the only known fermions 
the initial nucleon can decay into. Two possibilities can be realized here: 
$\Delta $B=$\Delta $L or $\Delta $B=$- \Delta $L (B and L are the total baryon and 
lepton numbers respectively). The first possibility would lead to the conservation 
of (B$-$L) and second to the processes that violate (B$-$L) by two units. 
The most stringent nucleon decay limits are experimentally established 
\cite{PDG,NNN99,Shiozawa} for nucleon decay modes where (B$-$L) is conserved such as 
$p \rightarrow e^{+} + \pi ^{0},p \rightarrow \bar {\nu } + K^{+},
p \rightarrow \mu ^{+} + K^{0}$, etc. Experimental non-observation of these 
decay modes in IMB, Fr\'{e}jus, Kamiokande, Soudan II, Super-K experiments, and other 
experiments \cite{PDG} excluded original SU(5), one-step-breaking SO(10), and 
SUSY extended SU(5) models \cite{Pati}. Also, recent Super-Kamiokande results 
\cite{Kajita} almost rule out SUSY-extended SO(10) proton decay model \cite{Pati}. 
It is important to notice that in all these models, as well as in the Standard 
Model, (B$-$L) is strictly conserved at perturbative and non-perturbative levels. 
A new generation of experiments with huge-mass detectors \cite{NNN99} is needed to 
continue to test the stability of nucleons with respect to the (B$-$L) 
conservation. We believe that the search for the processes with (B$-$L) 
non-conservation will also need to be pursued by the future experiments 
on the equal basis. 

Why (B$-$L) might not be conserved? Naively we would expect that (B$-$L) 
number be strongly violated: the number of neutrons in our laboratory 
samples is in excess of equal number of protons and electrons. However, most 
leptons in the universe likely exist as, yet undetected, relic $\nu $ and 
$\bar {\nu }$ radiation similar to cosmic microwave background radiation of 
photons. Thus, the conservation of (B$-$L) on a scale of the whole 
universe remains an open question.

In nucleon decay (with nucleon disappearance $\Delta $B=$-$1) 
non-conservation of (B$-$L) implies the existence of transitions of the 
type $N \rightarrow lepton + X$ with $\Delta $(B$-$L)=$-$2. The 
conservation of (B$-$L) in nucleon decays corresponds to 
$N \rightarrow antilepton + X$ transitions. If (B$-$L) can be violated by 
two units, it is natural to assume, as also follows from the Unification 
models \cite{Mohapatra80,GellMann}, that processes with $\mid \Delta $L $\mid$=2 and $\mid 
\Delta $B $\mid$=2 as well as the processes with $\Delta $B=$- \Delta $L are 
the components of the common physics coming from the same energy scale. 
Examples of the former two processes would be heavy Majorana neutrinos with 
$\mid \Delta $L $\mid$=2 transitions of $\nu \leftrightarrow \bar {\nu }$ and 
the transitions of \nnbar with $\mid \Delta$ B $\mid$ = 2; examples of the 
latter transitions could be processes $p \rightarrow \nu \nu e^{+}, n \rightarrow \nu
\nu \bar {\nu}, p \rightarrow \nu \pi^{+} \pi^{-}$ etc. In 
Unification models of SO(10) type, massive Majorana neutrinos with $\mid 
\Delta $L $\mid$=2 transitions violating (B$-$L) by two units can generate the 
masses of conventional neutrinos through the ``see-saw'' mechanism \cite{GellMann}. 
Thus, the explanation of the masses of neutrinos can be linked with (B$-$L) and 
B non-conservation.

Since 1973, when (B$-$L) non-conservation was first considered 
theoretically \cite{Salam}, it was discussed within the framework of Unification 
models in a number of theoretical papers \cite{Mohapatra80,Davidson,Wilczek,Pati}.  
In the left-right symmetric SO(10) Unification models the violation of (B$-$L) arises 
simultaneously at the same energy scale as the violation of left-right symmetry 
\cite{Mohapatra80,Davidson}. Thus, (B$-$L) non-conservation is related with the 
searches of the right-handed currents and $W_{R}$ vector bosons. Present experimental 
lower limits for $W_{R}$ mass are in a TeV range \cite{PDG}. If \nnbar transitions 
would be experimentally observed beyond the existing experimental limits
it will point out to the energy scale of the L-R restoration much higher than a TeV.

The \nnbar transitions live on an energy scale $\sim$10$^{5}$ GeV. This scale 
follows from the dimensional reason: indeed, the disappearance of 3 quarks and 
appearance of 3 anti-quarks in \nnbar can be described as a Feynman operator of 
dimension 9 (each of six fermions brings 
to Lagrangian the dimension of $m^{3/2})$, therefore \nnbar transition amplitude 
is proportional to $\it{\emph{M}}^{-5}$, where $\it{\emph{M}}$ is the energy scale 
of (B$-$L) violation. Similarly, (B$-$L) conserving proton decay of the type 
$p \rightarrow e^{+} \pi^{0}$ is described by the operator of dimension 6 and 
transition amplitude is proportional to M$^{-2}$, where M is the unification 
scale. If the unification energy scale of SU(5) $\sim $10$^{15} -$10$^{16}$ GeV 
would make $p \rightarrow e^{+} \pi^{0}$ observable, the energy scale corresponding 
to observable \nnbar transition should be in the range of 10$^{5} - $10$^{6}$ GeV. 

Could one expect a new physics at such ``low energy scale''? The concept of 
Great Desert introduced by SU(5) model \cite{Georgi} was very popular for more than 
two decades. According to this concept no new physics would appear between 
electro-weak scale (or SUSY energy scale in the modified concept) 
and Grand Unification scale of $\sim $ 10$^{15} - $10$^{16}$ GeV. Recently 
it has been realized \cite{Arkani} that unification of gravitational and gauge 
interactions might occur at much lower energy scale and in a way that 
preserves broken L-R gauge symmetry. Therefore, the energy scale of 
10$^{5} - $10$^{6}$ GeV might be a natural scale for existence of \nnbar
transitions.

Probably the most compelling reason for the existence of (B$-$L) 
non-conservation follows from the theoretical observation \cite{KRS85} 
that electroweak non-perturbative "sphaleron" mechanism in the early 
universe would \textit{erase} the observed BAU if (B$-$L) is globally conserved. 
Therefore, it is natural to assume that (B$-$L) \textit{non-conservation} 
takes place at the energies above the electro-weak scale. In this sense, experimental 
discovery of the nucleon decay into ''standard'' decay modes like $p \rightarrow 
e^{+} \pi^{0}$ or $p \rightarrow \bar {\nu} K^{0}$ with conservation of (B$-$L) would 
leave BAU unexplained.

As was pointed out by Gell-Mann and Pais in 1955 \cite{GellMannPais}, the only conservation 
law of nature that would forbid the \nnbar transition is the conservation of 
\textit{baryon number}. In 1970, \nnbar transition was considered by Kuzmin as a 
possible explanation of BAU \cite{Kuzmin}. In the 1980s, it was suggested by Glashow in 
the context of SU(5) models \cite{Glashow} and independently by Marshak and Mohapatra 
\cite{Mohapatra80} in the context of left-right symmetric models that \nnbar transition 
could lead to theoretical schemes complementary or alternative to those exploiting 
the (B$-$L) conserving proton decay mechanism. In particular, Marshak and Mohapatra 
pointed out that there is an intimate connection between a non-vanishing Majorana mass 
for neutrinos and a possibility of the $\Delta $(B$-$L) =$-$2 in \nnbar transition. 
The recent experimental indications \cite{neutrinos} for the existence of neutrino mass  
therefore strengthen the case for a new dedicated search for \nnbar oscillations.  

The question of the energy scale where (B$-$L) violation manifests itself is an 
experimental one. New experiments searching for \nnbar transitions might answer it.
Several recent theoretical works provide encouraging motivation here.
In the recent paper \cite{Babu} Babu and Mohapatra predicted 
in the context of a wide class of SO(10)-based left-right symmetric neutrino mass 
seesaw models the upper limit for the appearance of \nnbar transitions that is within 
the sensitivity range that can be reached by the next generation \nnbar search 
experiments. Nussinov and Shrock \cite{Shrock} have analyzed the \nnbar process within 
the scope of generic models with large extra dimensions in which standard model fields 
propagate and fermion wave functions have strong localization. They found that in these 
models proton decay can be suppressed but \nnbar transitions might occur at the levels 
not too far below the current experimental limits. Dvali and Gabadadze \cite{Dvali} 
discussed how global 
charges can be violated in the extra-dimensional models with low quantum gravity scale.
In their approach the Standard Model particles are localized on the brane, which is 
a fluctuating object. Due to quantum fluctuations, there are following virtual 
processes occur: brane gets curved locally and creates a bubble ("baby brane") which 
gets detached from the brane and goes into extra dimension where it effectively becomes 
a black hole, then reenters again on the brane and decays there. Baryon number can be 
violated inside the black hole due to quantum gravity effects. The baby brane can 
take away any particle with strictly zero gauge charge, such as a neutron, and return back 
any other combination of the same mass and spin, for instance, anti-neutron. On the 
brane this process will be seen as \nnbar transition. The same process cannot lead to 
the proton decay, since the charged particles cannot leave the brane, because the 
photon is localized there. The amplitude of the \nnbar transition depends on the parameters 
such as fundamental Planck mass, brane thickness etc. For reasonable values the effective 
operator comes out to be suppressed by the scale $\sim$10$^{5}$ GeV.

\section{Experimental searches for \nnbar}

Observation of \nnbar transitions would be the most spectacular manifestation of 
the new physics, if such physics exists at the energy scale around 10$^{5}$ GeV. 
Let's assume that the amplitude of \nnbar transition given by this energy scale is 
$\alpha$. Then , the transition probability of \nnbar for free neutrons can be found 
\cite{Mohapatra80} from the solution of time-dependent Schr\"{o}dinger equation~(\ref{eq:tdsch}):

\begin{equation}
i\hbar \frac{\partial \psi }{\partial t} = \left( {{\begin{array}{*{20}c}
 {E_n } \hfill & \alpha \hfill \\
 \alpha \hfill & {E_{\bar {n}}} \hfill \\
\end{array} }} \right)\mbox{ }\psi 
\label{eq:tdsch}
\end{equation}

where \textit{$\psi$} describes a 2-component neutron-antineutron state, $E_{n}$
and $E_{\bar {n}}$ are non-relativistic operators of total energy for neutron and 
antineutron respectively. Probability of neutron to antineutron transition in the 
absence of external fields different for neutrons and anti-neutrons can be then 
found from (\ref{eq:tdsch}) as  $P_{n \bar {n}} = (t / \tau_{n \bar {n}})^{2}$  where 
$\tau_{n \bar {n}} = \hbar / \alpha$ is a characteristic transition (oscillation) 
time. 

Previous experimental searches \cite{PDG} for \nnbar transitions have been made 
(a) in vacuum with free neutrons from the reactors and (b) with neutrons bound inside 
nuclei. Sensitivity or discovery potential ($D.P.$) in experiments with free neutrons 
can be defined as a product of a number of neutron flights per second and the average 
square of the neutron flight time in a free space or between two collisions with walls. 
In the state-of-art experiment \cite{ILL} performed by Heidelberg-ILL-Padova-Pavia 
Collaboration at HFR reactor with cold neutron beam at ILL/Grenoble the 
discovery potential of 1.5$\cdot$10$^{9}n \cdot s$ have been reached. After approximately 
a year of beam exposure it resulted in the limit for free-neutron \nnbar transition 
time of $\tau_{n \bar{n}}\geq 8.6\times 10^{7}$ \textit{seconds}.

In intranuclear transitions \nnbar probability is suppressed \cite{nuctheory} by the 
presence of nuclear potential in (\ref{eq:tdsch}) that is different for neutron and antineutron. 
This suppression under reasonable assumptions can be calculated by nuclear theory 
\cite{nuctheory} with an accuracy  \cite{Kopeliovich} of about factor of 2. Corresponding 
intranuclear lifetime is given by 

\begin{equation}
\textit{$\tau$}(\textit{intranuclear}) = R \cdot \tau_{n \bar{n}}^{2}(\textit{free}) 
\label{eq:intra}
\end{equation}

where R is dimensional "suppression factor" equal to $\sim 2\cdot$10$^{23}$\textit{sec}$^{-1}$ 
for the most of nuclei \cite{nuctheory}. The order of magnitude of intranuclear suppression can be
qualitatively understood from the following simple considerations. The neutron bound in nucleus 
by energy $\Delta E$ ($\sim$10 \textit{MeV}) is quasi-free for the time 
$\Delta t \sim 1/ \Delta E$ or $\approx $10$^{-22} sec$. Probability of \nnbar transition 
during the time $\Delta t$ is ($\Delta t/ \tau_{n \bar{n}}$)$^{2}$. This condition occurs 
$1/ \Delta t$ times per second. Then $\tau$ (\textit{intranuclear})=$\tau_{n \bar{n}}^{2}/ \Delta t$.
One can see that $1/ \Delta t \approx 10^{22}$ \textit{sec}$^{-1}$ plays the role or 
suppression factor $R$. This factor is different from more carefully calculated factors in
\cite{nuctheory} by only 1-2 orders of magnitude. 

Most stringent limit in the intranuclear \nnbar transition search was provided by 
Soudan 2 Collaboration \cite{Soudan2}. Their limit for iron nucleus lifetime
with respect to intranuclear \nnbar transition is $\tau_{Fe} \geq$ 7.2$\cdot$10$^{31}$ 
\textit{years}. This according to (\ref{eq:intra}) and intranuclear suppression factor
\cite{nuctheory} corresponds to free neutron oscillation time 
$\tau_{n \bar{n}}\geq 1.3\times 10^{8}$ \textit{seconds}. 
Intranuclear \nnbar search also is being pursued by Super-K \cite{Ganezer} and
SNO \cite{Formaggio} collaborations. It is expected \cite{Mann} that limits 
reached by both experiments will be in the range of 
$\tau_{n \bar{n}}\geq 4\times 10^{8}$ \textit{seconds}. However, since their
observations will be limited by the atmospheric neutrino background (0.7 $\nu$'s per 
kt-year in Soudan 2 experiment) further improvement in $\tau_{n \bar{n}}$
will be proportional to (kt-year)$^{1/4}$. Interesting that SNO sensitivity is close to 
that in Super-K although SNO detector mass is much smaller. This is mainly due to 
order-of-magnitude smaller nuclear suppression factor for deuterium \cite{nuctheory}.
 
From the comparison of free-neutron and bound-neutron techniques one can clearly see 
the advantages of the experiments with reactor neutrons: an order of magnitude improvement 
in the sensitivity of experiment with free neutrons is equivalent to two orders of magnitude 
improvement in intranuclear search. Presently both methods give approximately equal
limits.

Recent progress in increase of production of ultracold neutrons allows one to plan 
a search for \nnbar transitions with UCN \cite{UCNworkshop}. Sources of ultracold
neutrons are being constructed or planned at LANL, PSI/Zurich, FRMII/M\"{u}nchen, 
in Japan, at North Carolina State University reactor, and at Indiana University compact 
neutron source. The UCN production rate at some of these sources is expected to 
exceed 10$^{7}$\textit{s}$^{-1}$. With UCN production rate of 
$\sim$5$\cdot$10$^{7}$\textit{s}$^{-1}$ or higher and with assumption that 
detector can be made backgroundless it will be possible to reach 
the discovery potential $D.P.>$4$\cdot$10$^{10}n \cdot s$ and therefore search 
for \nnbar transition up to the limit of $\sim\tau_{n \bar{n}}\geq 10^{9} s$.

The maximum sensitivity for the \nnbar search is provided by use of intensive cold-neutron 
beam from the high-flux research reactors \cite{YK2}. With the presently available 
technology  (focusing neutron reflector and cold neutron moderator) and with existing 
sources of neutrons (e.g. HFIR reactor at ORNL with the world highest thermal flux 
of $\sim$1.5$\cdot$10$^{15}$ n$\cdot$cm$^{-2}\cdot$s$^{-1})$ it is possible \cite{YK2} 
to reach sensitivity $D.P.>$6$\cdot$10$^{11} n \cdot s$ and search for \nnbar 
transitions up to and above the limit of $\tau_{n \bar{n}}>$3$\cdot$10$^{9} s$.
This sensitivity (i.e. time integrated probability of antineutron appearance) is 
1,000 times higher than in Heidelberg-ILL-Padova-Pavia experiment\cite{ILL}. The 
experimental signature of appearance of antineutron in the cold-energy neutron beam 
would be unambiguous and background free as was demonstrated by the ILL/Grenoble-based 
experiment \cite{ILL}. The \nnbar oscillation time limit that could be reached in a new 
HFIR-based experiment would correspond to intranuclear lifetime limit of 
$\sim $ 10$^{35}$ years, which exceeds significantly the sensitivity for matter 
instability search in Super-K.

\begin{figure}
\begin{center}
\psfig{figure=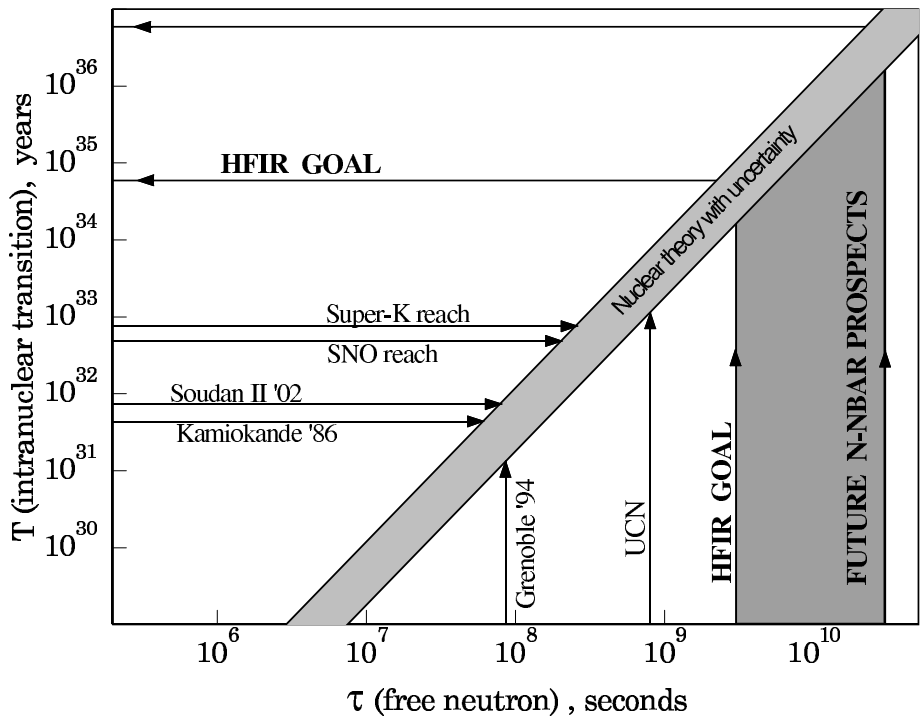,width=5.0in,clip}
\caption{Comparison of free-neutron and bound-neutron methods for 
        \nnbar transition search. Horizontal scale: limits for characteristic 
        transition time in experiments with free neutrons from the reactors. 
        Vertical scale: limits for lifetime for intranuclear \nnbar transition 
        in nucleon decay experiments. Two methods are related by theoretical 
        ``nuclear suppressed'' dependence [27] as explained in the text.}
\label{fig1}
\end{center}
\end{figure}

\begin{figure}
\begin{center}
\psfig{figure=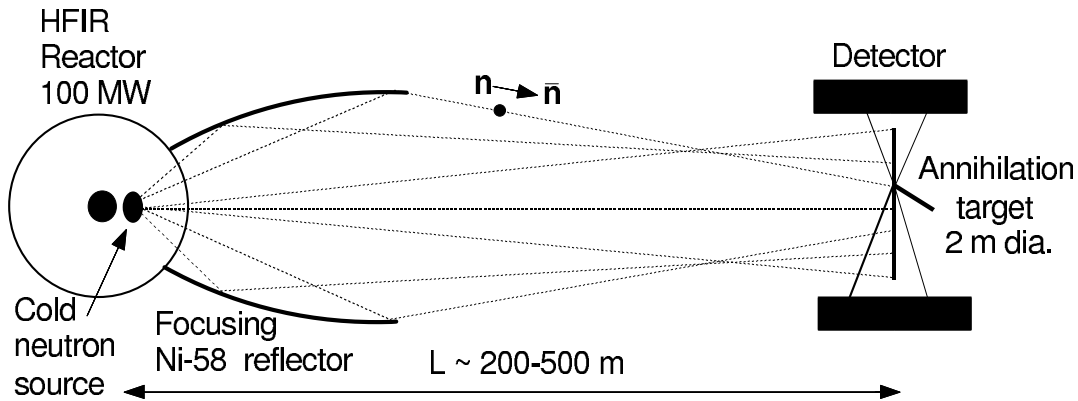,width=6.0in,clip}
\caption{Conceptual layout of cold-neutron beam \nnbar search experiment 
with focusing reflector for HFIR/ORNL research reactor (not to scale).}
\label{fig2}
\end{center}
\end{figure}

Present \nnbar limits and future sensitivity reaches for different experimental 
techniques (cold neutron beam, UCN, and intranuclear) are compared in Figure
\ref{fig1}. Conceptual layout of cold-neutron beam experiment at HFIR/ORNL reactor  
is shown in Figure \ref{fig2}.

\begin{figure}
\begin{center}
\psfig{figure=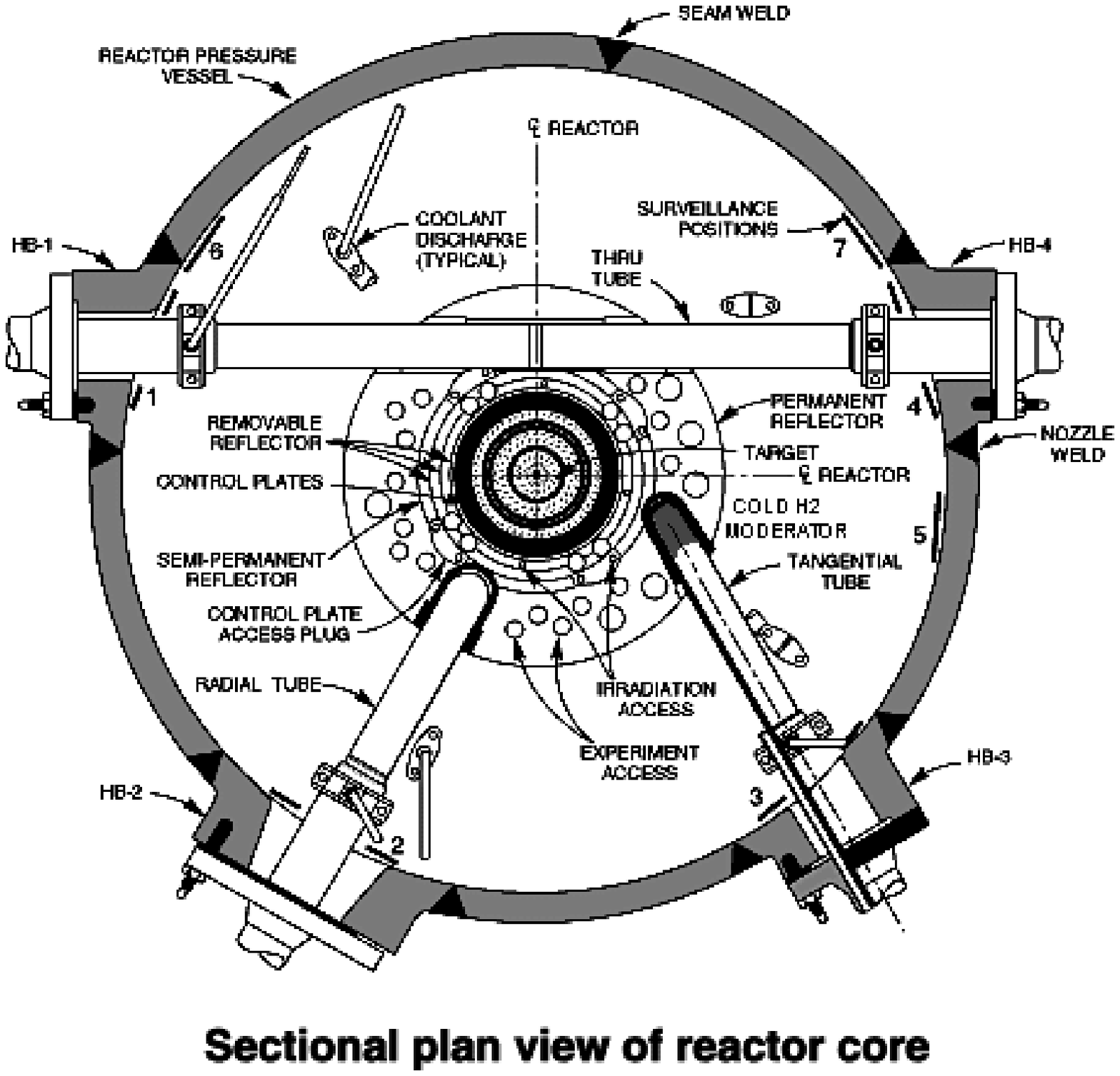,width=4.0in,clip}
\caption{Section view of ORNL/HFIR reactor core. In the \nnbar 
search experiment the cold supercritical hydrogen moderator should be 
installed in the HB-3 beam tube.}
\label{fig3}
\end{center}
\end{figure}

Figure \ref{fig3} shows cross-section of the HFIR reactor at ORNL where 
\nnbar search experiment can be implemented at HB-3 beam line equipped with 
a new cold neutron moderator.

One can show, ignoring the effect of gravity, that the probability of 
observation of antineutron in the experiment with optimized focusing 
reflector is proportional to $D.P. \sim L^{2 }/T^{3 / 2}$, where $L$ is the 
distance between the point of reflection and the annihilation detector and $T$ 
is an effective temperature of the thermalized neutron spectrum. This should be 
compared with the discovery potential for the layout without the focusing 
reflector where $D.P. \sim 1/T^{1 / 2}$. Thus, with the advanced layout the 
large length of the experiment and the low temperature of neutrons would result 
in substantial increase of the discovery potential. More comprehensive 
Monte-Carlo simulations including gravity effect \cite{YK2} show that discovery 
potential of HFIR-based experiment with cold supercritical hydrogen 
moderator \cite{HB4} in HB-3 beam pipe (see Figure \ref{fig3}) can be more than 
factor of $\sim $ 400 higher than in ILL/RHF-based experiment \cite{ILL}. 
Thus, \textit{one day} of operation at HFIR in the new proposed \nnbar search 
\cite{Bugg} is equivalent to \textit{one year} in the previous ILL/RHF-based 
experiment. Table \ref{tab1} compares essential features of 
the new-proposed HFIR experiment \cite{Bugg} in HB-3 beam port with the 
ILL/RHF-based experiment \cite{ILL}.

\begin{table}[htbp]
\caption{Comparison of the major parameters of a new \nnbar search experiment
proposed for HB-3 beam line at High Flux Isotope Reactor at Oak Ridge National
Laboratory with the previous \nnbar search experiment performed in 1989-1991
at RHF Reactor at ILL/Grenoble}
\begin{center}
\begin{tabular}
{|p{145pt}|p{117pt}|p{110pt}|}
\hline
Neutron source& 
RHF/Grenoble& 
HFIR/ORNL  \\
\hline
Reference& 
[26]& 
[35] \\
Status of experiment& 
Completed& 
Proposed (HB-3 beam) \\
Reactor power, MW& 
58& 
(85) 100 \\
Reactor's peak thermal n-flux & 
1.4 10$^{15}$ (n/cm$^{2}$/s)& 
1.5 10$^{15}$ (n/cm$^{2}$/s) \\
Moderator& 
Liquid D$_{2}$& 
Supercritical H$_{2}$ \\
Source area& 
6$\times$12 cm$^{2}$& 
$\sim$ 11 cm diameter \\
Target diameter& 
1.1 m& 
2.0 m \\
Flight path& 
76 m& 
300 m \\
Neutron fluence at target& 
1.25$\cdot$10$^{11}$ n/s& 
$\sim$ 8.5$\cdot$10$^{12}$n/s \\
Average time of flight& 
0.109 s& 
0.27 s \\
Detector efficiency& 
0.48& 
$\sim$ 0.5 \\
Operation time (s)& 
2.4$\cdot$10$^{7}$& 
7$\cdot$10$^{7}$ ($\sim $3 years) \\
Discovery potential per sec & 
1.5$\cdot$10$^{9}$ n$\cdot$s$^{2}$& 
6.2$\cdot$10$^{11}$ n$\cdot$s$^{2}$ \\
$\tau_{n \bar {n}}$ limit (90{\%}CL)& 
8.6$\cdot$10$^{7}$ s& 
3.0$\cdot$10$^{9}$ s \\
\hline
\end{tabular}
\label{tab1}
\end{center}
\end{table}

The conceptual scheme of the antineutron annihilation detector in the 
new experiment can be similar to that used in the previous Heidelberg-ILL-Padova-Pavia 
experiment \cite{ILL}. Annihilation target is a thin carbon-film membrane with almost 
100{\%} efficiency for antineutron detection and with low efficiency for (n,$\gamma$ ) 
conversion. Final states of nucleon-antinucleon annihilation are well understood mainly 
due to the LEAR studies and can be accurately modeled \cite{Golubeva}. Average final state 
has five pions originating in the annihilation target. Tracking part of the detector 
should reconstruct the candidate event vertex and verify its position relative the 
annihilation target origin. A calorimeter can be used for triggering and for the total 
energy deposit measurement (below $\sim $1.8 GeV). Detector should be surrounded by cosmic 
veto scintillator counter system to reduce the trigger rate and to remove the possible 
cosmic ray background. 

Quality of the $^{58}$Ni coating of the focusing reflector does not need to 
be as perfect as in the case of conventional neutron guide mirrors and 
super-mirrors since neutrons undergo essentially only single reflection. 
Vacuum in the flight tube should be better than 10$^{ - 4}$ Pa \cite{ILL}.
Earth magnetic field that would suppress the \nnbar transition must be 
compensated down to a level of few nano-Tesla in the entire flight volume. 
Following the recommendation of \cite{ILL} both active (compensating coils) 
and passive (permalloy) screens should be used to achieve required field 
compensation. An active magnetic field compensation system provides cross 
check by ``switching off'' the effect in the case if antineutron signature 
is observed. 

\section{Beyond the existence of $n \rightarrow \bar{n}$}

It was pointed out in paper \cite{Abov} that the existence of \nnbar transitions 
would provide a unique opportunity for testing CPT-theorem by looking at the 
mass difference $\Delta m$ of neutron and antineutron. Such mass difference 
(or similarly a small gravitational non-equivalence of neutron and antineutron) 
\cite{Lamoreaux} will suppress the \nnbar transition for free neutrons but will be 
too small to produce a sizable additional effect in intranuclear transitions 
where a very large suppression is already present due to significant difference 
of nuclear potentials for neutron and antineutron. Therefore, two measurements 
are required: one with free neutrons in the reactor experiment and the other 
with bound neutrons in intranuclear transitions \cite{Abov}. The latter experiment 
can be replaced by a reactor-based measurement with a controllable variation 
of the magnetic field. Since the ultimate sensitivity to $\Delta m$ of the 
reactor-based experiment is $\Delta m < \hbar/\Delta t$, with a neutron 
flight time $\Delta t \sim $ 0.3 sec (for HFIR-based experiment), the smallest 
achievable value of $\Delta m/m$ can be few orders of magnitude lower than
$m_{n}/m_{Plank}$.

\section{Conclusion}

Reactor search for \nnbar transition is a very sensitive method of 
observation of (B$-$L) non-conserving processes. Proposed new experimental 
approach to search for \nnbar transition at research reactors with cold 
neutron beams might result in an equivalent experimental limit of 10$^{35}$ 
years for intranuclear stability of nucleon. Such limit is not attainable by 
any other existing experimental method. 

Physics at the energy scale of 10$^{5} - $10$^{6}$ GeV might be dominated 
by quantum gravity effects that could be manifested through \nnbar transitions. 
If \nnbar transitions are observed, it will reveal a new phenomenon leading 
to a new physics at the energy scale beyond the range of colliders. New 
symmetry principles determining the history of the universe during the first 
moments of creation might be established; the left-right symmetry, broken in 
the Standard Model, may be found restored. The discovery of \nnbar transition 
would provide a major constraint on unification models and contribute to the 
understanding of baryon asymmetry of the universe. If and when such phenomenon 
is established, the subsequent experiments with \nnbar transition should allow 
a most precise test of CPT invariance and/or gravitational equivalence of 
baryonic matter and antimatter.

\end{document}